\documentclass{article}

\usepackage{amsmath}
\usepackage{enumitem}
\usepackage{quoting}
\usepackage{epigraph}
\usepackage{hyperref}

\usepackage{natbib}
\bibpunct{[}{]}{,}{a}{}{;}

\title{The Mysteries of Lisp -- I: \\ The Way to S-Expression Lisp}
\author{Hong-Yi Dai}
\date{\today}

\begin{document}
\maketitle

\begin{abstract}
 Despite its old age, Lisp remains mysterious to many of its admirers.  The
mysteries on one hand fascinate the language, on the other hand also obscure
it.  Following Stoyan but paying attention to what he has neglected or omitted,
in this first essay of a series intended to unravel these mysteries, we trace
the development of Lisp back to its origin, revealing how the language has
evolved into its nowadays look and feel.  The insights thus gained will not
only enhance existent understanding of the language but also inspires further
improvement of it.

\end{abstract}

\section{Introduction}\label{introduction}

\epigraph{You have to know the past to understand the present.}{\sc Carl Sagan}

\noindent
\citet{Graham2001Roots} uncovered ``the roots of Lisp'',\footnote{Throughout
this text, we will use the modern name Lisp for the language.  However, when
mentioning historical dialects or quoting text about them, we will stick to the
ancient name LISP.} and in particular, showed us ``the surprise''\footnote{The
one who first did so was of course \citet{McCarthy1959Recursive}.} that a
meta-circular interpreter for the language can be readily constructed.  This
surprising result, which he nominated as ``the defining quality of Lisp'', that
the language ``can be written in itself'' [ibid, p.~1], has attracted a lot of
language enthusiasts to Lisp.  However, when they ask about the source of this
surprise, they get answers such as that the language is Turing-complete or that
programs are (manipulable as) data in the language.  These answers, although
succinct, are nebulous.  To a large extent, the language remains mysterious.

We intend in a series of essays to unravel the mysteries of Lisp.  In the past
decades, some scholars have tried to do so and achieved the goal to various
degree.  The most remarkable is probably Stoyan who carefully studied the
history of Lisp for better understanding the language
\citeyearpar{Stoyan1979LISP, Stoyan1984Early, Stoyan1991Influence,
Stoyan2007Lisp, Stoyan2008Lisp}.  Holding the same position, we believe that to
unravel the mysteries of Lisp, we must trace its development back to its
origin.  So this work could be considered a continuation of that by Stoyan.  We
believe most of what we discuss here and will discuss in the following series
is folklore knowledge.  Moreover, it must have already been investigated by
Stoyan.  However, we intend not to repeat what Stoyan has done, but to
complement his work by gathering what he has neglected or omitted.  Our main
contribution is exhibiting our findings in one place and interpreting them in
both historical and modern contexts.

In this first essay of the series, we will look into the early development of
Lisp.  We will in particular lay out how the language has evolved into its
nowadays look and feel.  On just these aspects, a few pointers suffice to show
that our grasp of the language is not thorough: we may have heard that Lisp
became based on S-expressions more by accident, but have not considered any
less-accidental factor; we may have learned that lists are constructed from
pairs in Lisp, but not questioned the rationality of this actuality; we may
have read that Lisp used to have a kind of expressions other than S-expressions
for writing programs, but not investigated the context of their existence.

As far as we know, these issues have not, if ever, been satisfactorily
addressed.  A probable reason is that most of us never go beyond the landmark
paper \citep{McCarthy1960Recursive} that systematically described the language.
This paper, however, was not the first systematic description of Lisp.  Its
earlier draft \citep{McCarthy1959Recursive}, published as an AI memorandum,
was.  It is this AI memo on which we will focus our attention, and prior AI
memos in which we will seek useful clues.  In our opinion, McCarthy presented
in \citeyearpar{McCarthy1959Recursive} a better-designed system than the
later-determined version in \citeyearpar{McCarthy1960Recursive}, which
evidences again that the development of a system may not necessarily be an
advancement but rather a regression.

\section{Toward S-expression Lisp}\label{toward-s-expression-lisp}

In this section, we will look into the early development of Lisp, to a large
extent in chronological order.  We will focus our attention mainly on material
related to the three issues we set out to address.  For a more-general
treatment of the early history of Lisp, the reader is referred to either
\citet{Stoyan1979LISP, Stoyan1984Early, Stoyan1991Influence, Stoyan2007Lisp,
Stoyan2008Lisp} or \citet{McCarthy1978History, McCarthy1980LISP}.

\subsection{An algebraic language}\label{an-algebraic-language}

Although according to \citet{Stoyan1984Early}, the incubation of Lisp could
date back to 1956, it was in 1958 when the 1st AI memo was published that ``an
algebraic language for the manipulation of symbolic expressions'' (not named
LISP yet) \citep{McCarthy1958Algebraic} hatched out.  McCarthy put in his new
algebraic language most of the features he proposed ``for the Volume 2 (V2)''
of the International Algebraic Language\footnote{IAL, later known as ALGOL, the
ALGOrithmic Language} \citeyearpar{McCarthy1958Some}.  Among these features, of
particular interest to us is the proposal of an intermediate language that
uniforms function position: a function-designating expression, even an
operator, appears always in the head position, as in \(f\texttt{(}e_1\texttt{,}
\ldots\texttt{,} e_n\texttt{)}\).  Another intriguing feature is representing
both sequences and expressions as lists implemented using series of machine
words on IBM 704.  From this vivid description, we can already see a prototype
of Lisp.  However, note a few things.

First, list was introduced as a kind of \emph{data structure}, not yet
abstracted as a \emph{data type}.  We diverge from Stoyan on this point.
Stoyan held that ``lists were not regarded as data structures''
\citeyearpar[p.~304]{Stoyan1984Early}.  We believe they were, since ``a number
of interesting and useful operations on lists have been defined''
\citep[p.~5]{McCarthy1958Algebraic}.  They were used for constructing lists and
selecting components.  The omission of lists ``as a kind of
quantity''\footnote{In the context, the word `quantity' probably meant datum or
literal.  Taking into account the absence of symbolic expressions from the
algebraic language (discussed soon), this seems a plausible interpretation.}
was because ``most of the calculations we actually perform cannot as yet be
described in terms of these operations'' and ``it still seems to be necessary
to compute with the addresses of the elements of the lists'' [ibid, p. 5].
Although it is not clear what McCarthy meant by ``cannot as yet be described''
here, it is obvious that he felt that these operations were too low-level.

Second, symbolic expressions were not part of the algebraic language.  They
belonged in both the intermediate language and the language of discourse
(English plus mathematics).  McCarthy here changed for symbolic expressions
from the function notation \(f\texttt{(}e_1\texttt{,} \ldots\texttt{,}
e_n\texttt{)}\) in \citeyearpar{McCarthy1958Some} to the sequence notation
\(\texttt{(}f\texttt{,} e_1\texttt{,} \ldots\texttt{,} e_n\texttt{)}\), which
resembled the prefix notation except explicit parenthesization.

Third, the algebraic language itself used mixfix notation.  Although not
explicitly stated in \citep{McCarthy1958Algebraic}, according to the proposal
in \citep{McCarthy1958Some}, programs in the algebraic language were supposed
to be translated into the intermediate language and then further translated
into the assembly language or machine language.  Thus the first-stage
translation will turn algebraic expressions in mixfix notation completely into
symbolic expressions in prefix notation.

The algebraic language received two revisions documented respectively in AIM-3
\citep{McCarthy1958Symbol} and AIM-4 \citep{McCarthy1958Manipulating}.  AIM-3
made explicit that IBM-704 word sequences were internal representations of
algebraic expressions while symbolic expressions external.  Following the
distinction was an attempt to formally define symbolic expressions,
spontaneously called ``external expressions'':\footnote{Our edits are put in
square brackets.}
\begin{quote}
\begin{enumerate}[noitemsep,nolistsep]
\item
  A symbol is an [external] expression.
\item
  If \(e_1\), \(e_2\), \ldots, \(e_n\) are [external] expressions, so is
  \(\texttt{(}e_1\texttt{,} e_2\texttt{,} \ldots\texttt{,} e_n\texttt{)}\).
\end{enumerate}
\end{quote}
McCarthy further noted the distinction between \(e\) and
\(\texttt{(}e\texttt{)}\).  But the special case for empty sequence was
missing.  Still, symbolic expressions were not admitted into the algebraic
language.  However, operations on lists were distilled into more or less
\texttt{cons}, \texttt{car} and \texttt{cdr}.  In AIM-4, the name LISP, for
``List Processor'' \citep[p.~9]{McCarthy1958Manipulating}, first occurred.

\subsection{A list processor}\label{an-list-processor}

In the 8th AI memo \citep{McCarthy1959Recursive}, the draft of the landmark
paper \citep{McCarthy1960Recursive}, symbolic expressions finally entered the
algebraic language.  Lisp began to feature two systems of notation at the
source level: \emph{S-expressions} (short for Symbolic expressions) and
\emph{F-expressions}\footnote{`F-expression'
\citep[p.~13]{McCarthy1959Recursive} was the original name of M-expressions
(for Meta-expressions) \citep[p.~187]{McCarthy1960Recursive}.  We choose the
old name for two reasons: (1) \emph{meta} is relative, the name M-expression
becomes misleading once F-expressions are taken to the level of object
language; (2) the term F-expression clearly indicates that functional
expressions describe functions of S-expressions, which McCarthy called
S-functions \citeyearpar[p.~1]{McCarthy1959Recursive}.  Nevertheless, in
directly quoted text, we will keep the terms used in the source.} (for
Functional expressions), which respectively correspond to the source-level
forms of \emph{data} and \emph{programs}.

In this memo, McCarthy presented a system different from what was later given
in \citeyearpar{McCarthy1960Recursive}, and thus even different from what we
know today.  The definition of S-expressions
\citep[p.~3]{McCarthy1959Recursive}, now complete, is quoted below:\footnote{We
have slightly edited the original definition to fit modern typographic style.
In particular, we use \texttt{()} in place of the hand-written
\(\mathtt{\Lambda}\) for the null expression.  Again, our edits are put in
square brackets.}
\begin{quote}
\begin{enumerate}[noitemsep,nolistsep]
\item
  The atomic symbols [...] are S-expressions.
\item
  A null expression [\texttt{()}] is also admitted.
\item
  If \(e\) is an S-expression[,] so is \(\texttt{(}e\texttt{)}\).
\item
  If \(e_1\) and \(\texttt{(}[e_s]\texttt{)}\) are S-expressions[,] so is
  \(\texttt{(}e_1\texttt{,} [e_s]\texttt{)}\).
\end{enumerate}
\end{quote}
\(e_s\) in the 4th clause might refer to a sequence of S-expressions
``\(e_2\texttt{,} \ldots\texttt{,} e_n\)'' where \(n \ge 2\).  In that case,
the rule gives us \(\texttt{(}e_1\texttt{,} e_2\texttt{,} \ldots\texttt{,}
e_n\texttt{)}\).  This notation for lists is almost the same as what we know
today except that it used commas rather than merely spaces to separate list
elements.

According to this definition, valid compound S-expressions include only what we
now call \emph{proper lists} that always terminate with \texttt{()}, \emph{no}
ordered pairs, and naturally \emph{nor} improper lists that do not end with
\texttt{()}.\footnote{From now on, when the word `list' occurs without any
qualifier, it means proper list, as has already been the case.}

As regards the notation for functional expressions, McCarthy switched to square
brackets and semi-colons ``since parentheses and commas have been preempted''
by S-expressions \citeyearpar[p.~3]{McCarthy1959Recursive}.  This switch was a
move away from the familiar mathematical notation used since AIM-1, where
\(\texttt{f(}e_1\texttt{,} \ldots\texttt{,} e_n\texttt{)}\) were used for
functional expressions, and \(\texttt{(}e_1\texttt{,} \ldots\texttt{,}
e_n\texttt{)}\) for symbolic expressions.\footnote{We could not see how this
notation might cause any serious problem.}  Had McCarthy retained the notation
in AIM-1 or reversed the notation in AIM-8 for S-expressions and
F-expressions\footnote{Interestingly, a later Lisp dialect called M-LISP, which
his inventor advertised as a ``hybrid of McCarthy's original M-expression LISP
and Scheme'', did reverse the notation for S-expressions and F-expressions
\citep{Muller1991M-LISP, Muller1992M-LISP}.} --- in other words, if he used
square brackets and semi-colons for S-expressions while parentheses and commas
for F-expressions, that is, \(\texttt{[}e_1\texttt{;} \ldots\texttt{;}
e_n\texttt{]}\) instead of \(\texttt{(}e_1\texttt{,} \ldots\texttt{,}
e_n\texttt{)}\), and \(\texttt{f(}e_1\texttt{,} \ldots\texttt{,}
e_n\texttt{)}\) rather than \(\texttt{f[}e_1\texttt{;} \ldots\texttt{;}
e_n\texttt{]}\) --- Lisp would have a more-mathematical flavor, which would in
turn better justify its being roughly a language for ``a mathematical theory of
computation'' \citeyearpar{McCarthy1961Basis} based on ``recursive functions of
symbolic expressions'' \citeyearpar{McCarthy1959Recursive} and surely ``an
algebraic language for the manipulation of symbolic expressions''
\citeyearpar{McCarthy1958Algebraic}.

As McCarthy noted in the abstract, AIM-8 contained ``only the
machine[-]independent parts of the system''.  We see for the first time lists
be presented without mentioning memory addresses.  In other words, in AIM-8,
list got abstracted as a data type.  Operations on lists were defined by
axioms.  The definitions of the selector functions (called \texttt{first} and
\texttt{rest}\footnote{These two names were reintroduced by the PLT people
(\url{http://racket-lang.org/people.html}) into their variant of Scheme (now
called Racket, \url{http://racket-lang.org/}) for \texttt{car} and \texttt{cdr}
constrained (by their contract system) to accepting only proper lists as valid
input.} rather than \texttt{car} and \texttt{cdr}) and the constructor function
(called \texttt{combine} instead of \texttt{cons})
\citep[pp.~3--4]{McCarthy1959Recursive}, are reproduced below:
\begin{align*}
 \texttt{first}\texttt{[}\texttt{(}e\texttt{)}\texttt{]}
  & = e \\
 \texttt{first}\texttt{[}\texttt{(}e_1\texttt{,} e_s\texttt{)}\texttt{]}
  & = e_1 \\
 \texttt{rest}\texttt{[}\texttt{(}e\texttt{)}\texttt{]}
  & = \texttt{(}\texttt{)} \\
 \texttt{rest}\texttt{[}\texttt{(}e_1\texttt{,} e_s\texttt{)}\texttt{]}
  & = \texttt{(}e_s\texttt{)} \\
 \texttt{combine}\texttt{[}e\texttt{;} \texttt{(}\texttt{)}\texttt{]}
  & = \texttt{(}e\texttt{)} \\
 \texttt{combine}\texttt{[}e_1\texttt{;} \texttt{(}e_s\texttt{)}\texttt{]}
  & = \texttt{(}e_1\texttt{,} e_s \texttt{)}
\end{align*}
where \(e_s\) might be ``\(e_2\texttt{,} \ldots\texttt{,} e_n\)'' for \(n \ge
2\).  McCarthy further noted that \texttt{first} and \texttt{rest} are defined
only for S-expressions ``which are neither null nor atomic'', and that
\texttt{combine} is defined when \(e_s\) is not atomic.  Note that the
S-expression \texttt{()}, which represents a null list, was \emph{not}
considered an atomic symbol.  The constraint on the second argument of
\texttt{combine} prevents the construction of pairs, and in turn improper
lists.

Renaming the selector functions shows that McCarthy ``felt uneasy with the
machine[-]dependent names'' \citep[p.~416]{Stoyan1991Influence} already in use
since \citeyearpar{McCarthy1958Algebraic}.  Constraining the the second
argument of the constructor function to lists suggests that he probably
recognized the possible misuse of the too-liberal constructor function to build
improper lists.

The attempt to rename the selector functions failed and McCarthy reverted to
the cryptic names \texttt{car} and \texttt{cdr}, as ``the LISP community was
already more powerful [than] the designer''
\citep[p.~416]{Stoyan1991Influence}.  The attempt to constrain the constructor
function, was also abandoned, in our opinion, for no good reason as well.

\subsection{A symmetric variant}\label{a-symmetric-variant}

At the end of AIM-8, McCarthy proposed ``binary Lisp'' which was a variant that
admits ``only two[-]element lists'' \citeyearpar[p.~17]{McCarthy1959Recursive}.
In particular, the status of \texttt{first} and \texttt{rest} was symmetric by
definition:
\begin{quoting}[vskip=0em]
\setlength{\abovedisplayskip}{0em}
\begin{align*}
 \texttt{first}\texttt{[}\texttt{(}e_1\texttt{,} e_2\texttt{)}\texttt{]}
    & = e_1 \\
 \texttt{rest}\texttt{[}\texttt{(}e_1\texttt{,} e_2\texttt{)}\texttt{]}
    & = e_2 \\
 \texttt{combine}\texttt{[}e_1\texttt{;} e_2\texttt{]}
    & = \texttt{(}e_1\texttt{,} e_2\texttt{)}
\end{align*}
\setlength{\belowdisplayskip}{0em}
\end{quoting}
What this definition would give is surely \emph{not} ``two-element lists''
according to the definition given earlier, and had better be called pairs.
McCarthy obviously abused the notation for lists here.

Right below the presentation of this definition, McCarthy remarked, in this
binary variant, that only two predicates \texttt{=} (symbol equality) and
\texttt{atom} are needed (\texttt{null} dropped), and that ``the null list can
be dispensed with'' [ibid, p.~17].  This should not be interpreted as that he
tried to ditch the notion of null list because a mathematician knows well the
importance of \emph{null} and in \citeyearpar{McCarthy1960Recursive} he
introduced \texttt{NIL} exactly for it.  The correct interpretation is that he
proposed eliminating \texttt{()} as a separate case in the definition of
S-expressions, and treating it simply as an atomic symbol.  McCarthy further
pointed out that ``the system is easier until we try to represent functions by
expressions [\ldots]'' [ibid, p.17].  Here, he probably meant that the system
would lose its easy feel to the verbose nesting of pairs for building lists to
represent F-expressions.

The system given in \citep{McCarthy1960Recursive} turned out to be exactly this
binary variant, albeit reverted to the function names \texttt{car},
\texttt{cdr} and \texttt{cons}.  There McCarthy presented the simplified
definition of S-expressions [ibid, p.~187] as we know today:
\begin{quote}
\begin{enumerate}[noitemsep,nolistsep]
\item
  Atomic symbols are S-expressions.
\item
  If \(e_1\) and \(e_2\) are S-expressions, so is
  \(\texttt{(}e_1 \texttt{.} e_2\texttt{)}\).
\end{enumerate}
\end{quote}
He also resolved all the issues regarding binary Lisp as we see now and he saw
then.  The two components of a pair was separated by a dot instead of a comma.
The atomic symbol \texttt{NIL} was chosen to mark the end of a list.  The list
notation \(\texttt{(}e_1\texttt{,} e_2\texttt{,} \ldots\texttt{,}
e_n\texttt{)}\) was defined as syntactic sugar for \(\texttt{(}e_1 \texttt{.}
\texttt{(}e_2 \texttt{.} \texttt{(}\ldots\texttt{(}e_n \texttt{.}
\texttt{NIL}\texttt{)}\ldots\texttt{)}\texttt{)}\texttt{)}\).

The system indeed feels simpler.  However, it also exposes the underlying
representation of lists.  Naturally, the constraint on the constructor function
of lists was abandoned so as to allow the construction of pairs, and the
unconstrained \texttt{cons} which mirrors the blind behavior of the
corresponding machine instruction returned.  As a consequence, improper lists
found their way back.  Every function expecting a list as argument now should
test the argument to see if it is indeed proper, otherwise, it would fail when
it receives instead an improper list or a circular list.  But given that
improper lists are seldom used and the test usually has a linear-time
complexity, Lisp programmers either leave it out and assume the input to be a
proper list by wishful thinking, or treat the last \texttt{cdr} as \texttt{NIL}
in the case of an improper list and let the trap into an endless loop open in
the case of a circular list.

\subsection{F-expressions vs. S-expressions}\label{f-expressions-vs.s-expressions}

McCarthy himself always preferred and expected to ``[write] programs as
M-expressions''\footnote{McCarthy's expectation was fulfilled in the
short-lived LISP 2 \citep{Abrahams1966LISP}.}
\citeyearpar[p.~179]{McCarthy1978History}.  Before the first implementation of
Lisp came out, programs were indeed written as F-expressions and then
hand-compiled to assembly code.\footnote{The assembly language was SAP
(Symbolic Assembly Programs) for IBM 704.}  These two facts give us a good
reason to believe that the implementation McCarthy expected was a compiler that
compiles Lisp programs directly or indirectly via S-expressions to assembly
code.  Our belief is also supported by McCarthy's own words.  His remark ``you
are confusing theory with practice'' \citep[p.~307]{Stoyan1984Early} on
Russell's proposal of programming the universal function\footnote{The universal
function, \texttt{eval} according to \citet{McCarthy1978History} but
\texttt{apply} according to \citet{Stoyan2008Lisp}, was an F-expression.} in an
assembly language by hand\footnote{For other possible versions of the story,
see \citep{Stoyan2008Lisp}.  Whatever version, the story shows how difficult
but also how important it is for theoreticians and practitioners to
communicate.} suggests that, at that time he was not immediately aware that
Russell had proposed an implementation of Lisp by interpreting intermediate
program representations in the form of S-expressions.  Under Russell's
proposal, once source programs (in the form of F-expressions) are translated
into intermediate programs (completely S-expressions) following the rules of
translation first described informally in \citep{McCarthy1959Recursive} and
later specified formally in \citep{McCarthy1960Recursive}, the hand-compiled
implementation of the universal function could readily interpret them.
McCarthy later did realize that what Russell obtained by hand-compilation of
the universal function ``certainly was'' a Lisp interpreter
\citep[p.~307]{Stoyan1984Early}.

What followed, which was probably one of the most dramatic events in the
history of programming languages, that early adopters of Lisp went ahead
programming in the intermediate language of S-expressions rather than the
source language of F-expressions, was totally against McCarthy's expectation!
There might be technical reasons (for example, the translation scheme from
F-expressions to S-expressions was not implemented yet) or historical factors
(for instance, Russell advertised his result as an interpreter for Lisp source
programs) for the incident.  However, we believe the crucial reason is that the
S-expression language (S-language), although used as intermediate language, was
high-level enough for programming and even facilitating program construction.
Indeed, the S-language is as high-level as the F-language (of F-expressions),
since what gets changed through the ``trivial''
\citep[p.~2]{McCarthy1959Recursive} translation is only notation, \emph{not}
abstraction as in later-developed systems that translate source programs in
some high-level language to some low-level intermediate language like the JVM
byte code \citep{Lindholm1999Java} or the LLVM intermediate representation
\citep{Lattner2004LLVM}.

\section{Conclusion}\label{conclusion}

This investigation into the early development of Lisp shows how the look and
feel of the language was shaped by mathematics and mechanics.  The language was
designed for writing programs algebraically.  The algebraic feel was reflected
in functional expressions (or F-expressions).  The intention to represent them,
internally in the machine led to the introduction of the list data structure,
and externally in an intermediate language to the invention of symbolic
expressions (or S-expressions).  Gradually, list got abstracted as a data type
and S-expressions admitted into the source-level language.\footnote{The
rationale for this admission will be covered in the second essay of this
series.}  After the first interpreter-based implementation of Lisp was running,
S-expressions wan out as the preferred language for programming.  In addition
to technical and historical reasons, the incident could also be credited to the
identical abstractive power of S-expressions with F-expressions.

The design presented in \citeyearpar{McCarthy1959Recursive} suggests that the
very basic compound data type McCarthy wanted to include into Lisp was list,
\emph{not} pair.  This suggestion was justified by McCarthy's adherence to the
mathematical notion of sequence in AIM-1 through AIM-8.  After all, to process
lists was one of the design goals of the language.  Moreover, two-element lists
cover all possible use cases of pairs.  Some people may try to defend the
status of pairs by appealing to space efficiency (since when storing two
elements, a pair uses one less \texttt{cons}-cell than a list) or to obscure
data structures (such as circular lists that represent infinitely-repeated
sequences).  However, now that space is no longer a big issue and the
functionality of circular lists can be simulated by non-circular ones with a
loop (more precisely, with a jump back to the start at the end), the existence
of pairs in the language has become obsolete.

Examining the origin of the binary variant, we sense a mathematician's
commitment to symmetry and reductionism, which is yet another ``influence of
the designer on the design'' but which \citet{Stoyan1991Influence} has probably
overlooked.  The rationale McCarthy explicitly gave for proposing the binary
variant was that ``the unsymmetrical status of \texttt{first} and \texttt{rest}
may be a source of uneasiness'' \citeyearpar[p.~17]{McCarthy1959Recursive}.
The one he implicitly held was naturally for simplifying the system by reducing
lists to nested pairs.  However, both rationales were weakened by the resultant
system.  The symmetry was never used.  Instead, the asymmetry he tried to
eliminate was reintroduced, not by constraint but by convention.  The reduction
led to the dilemma we have seen, where the programmer either does \emph{nothing
sane} to bear it or \emph{something insane} to circumvent it.  We believe a
well-designed language should never put the programmer in such an awkward
situation.  One may propose including another set of manipulation functions,
say \texttt{first}, \texttt{rest} and \texttt{combine} as defined in the
earlier part of \citep{McCarthy1959Recursive}, particularly for lists.
However, it does not really solve the problem, only further complicates the
system.  If improper lists and circular lists are rarely used and can always be
simulated in the rare case, it is better to kick them out to favor the common
case.  We urge designers of new Lisp dialects to discard pairs and return to
lists as presented in \citep{McCarthy1959Recursive}.

This concludes the essay.  In the following series, we will reveal other
mysterious aspects of Lisp such as the relationship between code and data, the
existence of a meta-circular interpreter, etc.  Lisp, as the first language
that embraces and integrates ideas from three major theoretical bases of
computation, namely Turing machine, lambda calculus and recursion theory, is a
goldmine worth deep digging.

\section{Acknowledgments}\label{acknowledgments}

We appreciate Herbert Stoyan's donation of his collection on Lisp and AI to the
Computer History Museum \citep{CHM2010Herbert}.  We especially thank Paul
McJones of the Computer History Museum's Software Preservation Group team, as
well as many others, for collecting, preserving and presenting historical
materials about Lisp \citep{McJones2005History}.

\bibliographystyle{plainnat}
\bibliography{references}
\end{document}